\documentclass[epj]{svjour}
\usepackage{graphicx}

\usepackage{amsmath,amssymb}

\usepackage[enableskew]{youngtab} 

\def\be{\begin{equation}}
\def\ee{\end{equation}}
\def\j{j'}

\def\be{\begin{equation}}
\def\ee{\end{equation}}

\begin{document}
\title{Entanglement of one-magnon Schur-Weyl states}
\author{Pawel Jakubczyk\inst{1} \and Yevgen Kravets\inst{2} \and Dorota Jakubczyk\inst{3}%
}                     
%
%
\institute{University of Rzesz\'ow, Institute of Physics, ul. Rejtana 16c, 35-959 Rzesz\'ow, Poland 
\and University of Strathclyde, Department of Physics, 16 Richmond Street, Glasgow G1 1XQ, Scotland, United Kingdom 
\and Rzesz\'ow University of Technology, The Faculty of Mathematics and Applied Physics, W. Pola 2, 35-059, Rzesz\'ow, Poland}
\date{Received: date / Revised version: date}
%
\abstract{
We investigate the entanglement properties of symmetry states of the Schur-Weyl duality. Our approach based on reduced two-qubit density matrices, and concurrence as the measure of entanglement. We show that all kinds of "entangled graphs", which describe the entanglement structure in Schur-Weyl states are completely coded in the corresponding Young tableau.
} 
\maketitle
\section{Introduction}
\label{intro}

Entanglement is one of the most important aspects of quantum information processing (see, for example \cite{nielsen_chuang,schumacher,divincenzo}). The theory of separability between two-qubit systems was introduced by Peres \cite{peres1} and Horodecki et. al. \cite{horodecki1} while the measure of bipartite entanglement has been proposed, for instance in Refs. \cite{wooters1,peres2,horodecki2}.
When we intend to generalise the notion of bipartite entanglement to multipartite systems one encounters many problems. They steam from the fact that quantum entanglement cannot be shared freely among many qubits \cite{coffman,connor}, what means that entanglement structure is imposed by a choice of quantum state, which creates some kinds of quantum channels between qubits within the system. These channels can be used to transmission information between qubits in quantum information processing, by use of combined action of local unitary operators assigned to each qubit \cite{buzek_complete_graf}.

There are many approaches to identify different kinds of multipartite entanglement (see, for example \cite{thapliyal,bennett,sabin1}). 
We intend to mention only one of them, described in the work of D\"{u}r \cite{dur1}. 
He introduced the notion of \emph{entangled molecules}, by considering  bipartite aspects of multipartite entanglement. He has shown that each mixed state of a multiqubit system generates the web of quantum channels which, if presented graphically, resembles the chemical molecules. 
This approach has been completed in the work of Plesch and Bu\v{z}ek \cite{plesch1}, they focused only on pure states of the system and showed that every such state is classified by the so called \emph{entangled graph}, a generalized version of entangled molecule. 
This method allows us to study the structure of multipartite entanglement in terms of bipartite reduced density matrices and concurrence as a quantitative measure of entanglement.

In the present paper we consider a class of states which are some specification of states analysed in the work \cite{dur1}, and 
point out that they have a simple combinatoric interpretation within the Schur-Weyl duality \cite{schur,weyl} applied to the Heisenberg magnet \cite{nasz1,nasz2}. We expect that this model can play an essential role in a quantum-mechanical description of the kinematics of multipartite states. It is well adopted to the Schur-Weyl duality scheme since the arena for kinematics of the model spans exactly the whole tensor power space $h^{\otimes N}$, where $h$ is the space of a single qubit. Thus, it serves as a good example of the system in which entangled graphs ''live''.
 In order to describe these states in detail, we inwoke the famous Schur-Weyl duality between the actions $A:\Sigma_N \times h^{\otimes N} \rightarrow h^{\otimes N}$ and $B: U(2) \times h^{\otimes N} \rightarrow h^{\otimes N}$ of the symmetric group $\Sigma_N$ on the set $\tilde N = \{ 1,2, \ldots, N \}$ of $N$ constituent identical objects, and the unitary group $U(2)$ of each object, respectively, on the tensor power space $h^{\otimes N}$, supposed to carry all quantum states of the composite system (cf. \cite{l3,l4} for a more recent compact presentation). Such approach generates a new set of basis elements, so called Kostka matrices at the level of bases \cite{jucys_murphy}, of the Hilbert space $\mathcal H$ of the magnet, which is well adapted to the symmetry. 
The fact that the two dual actions mutually commute, i.e. $[A,B]=0$, implies that the Schur-Weyl states can be classified by two quantum numbers $t,y$, where $t$ denotes semistandard Weyl tableau - basis elements of irreducible representation of the unitary group $U(2)$ and $y$ standard Young tableau - basis elements of irreducible representation of the symmetric group $\Sigma_N$ \cite{young_tableux,sagan}. In other words this is a generalisation of the total angular momentum basis that is useful for exploiting symmetry under permutations or collective unitary rotations in one dimensional spin systems. There is a bijection between the set of all magnetic configurations and appropriate irreducible base of the duality of Schur-Weyl, known as the Robinson-Schensted (RS) algorithm, which ascribes to each magnetic configuration $f$ a pair $P(f)$ and $Q(f)$ of Weyl and Young tableaux \cite{robinson,schensted,knuth}.

This duality can be applied within the Heisenberg model of the magnetic ring in several ways. In our approach we select from the magnetic ring an abstract system, defined by its space $\mathcal H^{1} =\mathbb C^{N}$, which includes all those quantum states of the magnet which correspond to a single spin deviation (particle is localised at a node of the magnet), from the ferromagnetic vacuum $|++ \ldots + \rangle$.  Such states form a basis which spans a space of all quantum states of so called magnonic qudit \cite{magnonic_qudit}. In this simple case, states are uniquely labelled only by standard Young tableau $y$ (the Weyl tableau $t$ is unnecessary in this case) (c.f. sections \ref{roz2} for details). In the sequel we call them \emph{the one-magnon Schur-Weyl states} and only such states we will consider here.

The aim of the present paper is to show that one-magnon Schur-Weyl states on the ring $\tilde N$ have a natural interpretation in terms of quantum entanglement, in a spirit  of those considered by D\"{u}r in \cite{dur1}.
We show that such a strictly combinatorial object as the Young tableau $y$ fully describes the structure and the values of entanglement in the system, and it is also shown that this object completely defines the entangled graph of the system prepared in the quantum state labelled by this tableau. We also show that structure of the entangled graph can be easily interpreted in terms of RS algorithm.

The paper is organised as follows. We start in Sect. 2 with a brief description of the Heisenberg magnet, next we introduce the irreducible basis states of the Schur-Weyl duality and show that they are labelled by Young tableaux. In Sect. 3 we consider the notion of entangled graphs and describe the structure of these graphs for the system which is in the states labelled by tableaux $y$. Section 4 is devoted to combinatorial aspects of entangled graphs in the language of RS algorithm.
In Sec. 5 we presents an example and conclude in Sect. 6 with summary of our results.

\section{Schur-Weyl states for a single spin deviation}\label{roz2}

Let us consider a one dimensional isotropic Heisenberg magnet of $N$ nodes, with single node spin $s=1/2$ and a symmetry under collective unitary \emph{rotations} (the action $B$) and permutations of subsystems (the action $A$).
In the language of quantum computation we can say that there is a system of $N$ qubits, each with the local basis 
$|i\rangle, i \in \tilde{2}=\{ 0,1\}$, where $0$ denotes spin up and $1$ spin down,
such that
\be\label{r3}
h = lc_{\mathbb{C}} \tilde 2, \,\,\, dim \, h = 2,
\ee
is the linear closure of the set $\tilde 2$ over the field $\mathbb C$ of complex numbers.
A natural product state of the whole system can be presented as
\be
|f\rangle = | i_1, i_2, ..., i_N \rangle,  \,\,\, i_j \in \tilde 2,\;  j \in \tilde N.
\ee
We call this state \emph{a magnetic configuration}. The set of all magnetic configurations $\tilde 2^{\tilde N}$ forms the computational basis 
\be\label{baza}
b= \big \{ |f\rangle \;\big |\; f \in \tilde 2^{\tilde N}\big \},
\ee
which spans the Hilbert space
$
\mathcal{H} = lc_{\mathbb{C}} \,\, b \cong h^{\otimes N}
$
of the system.

The space $\mathcal{H}$ of all quantum states, with $\mbox{dim} \; \mathcal{H} =2^N$, decomposes under the action $A$ of the symmetric group $A: \Sigma_N \times \mathcal{H} \rightarrow \mathcal{H}$ as
\be\label{hilbert}
\mathcal{H}=\sum_{r=0}^{N}\oplus \,\,\mathcal{H}^{r},  \;\;\; \mbox {dim} \; \mathcal{H}^r={N \choose r},
\ee
into subspaces $\mathcal{H}^r $, with the fixed number $r$ of \textit{Bethe pseudoparticles} (spin deviations), since the quantum number $r$ is invariant under the action $A$.
Basis states of a subspace $\mathcal{H}^r $ form an orbit $\mathcal{O}_r$ of the action $A$ of the group $\Sigma_N$ on the set $b$. 
Such an orbit carries the transitive representation $R^{\Sigma_N : \Sigma^{(N-r,r)}}$ with stabiliser $\Sigma^{(N-r,r)}$ being the Young subgroup of the form $\Sigma^{(N-r,r)} = \Sigma_{N-r} \times \Sigma_{r} $.
This transitive representation decomposes into irreducibles (irreps) $\Delta^{\lambda}$ of $\Sigma_N$
\be\label{roz_tranzytywnej}
R^{\Sigma_N : \Sigma^{(N-r,r)}} \cong \sum_{ \lambda \unrhd (N-r,r)} K_{\lambda \, (N-r,r)} \,\, \Delta^{\lambda},
\ee
where $K_{\lambda \, (N-r,r)}$ denotes the Kostka number i.e. the number of all semistandard Weyl tableaux of the shape $\lambda$ and the weight $(N-r,r)$, and the sum runs over all partitions $\lambda$ greater or equal to partition $(N-r,r)$ in dominance order \cite{kerber}.

In the following, in order to obtain special class of states, we restrict ourselves to the case of Heisenberg magnet with only one spin deviation i.e. $r=1$. 
In this case the Hilbert space of such model is $\mathcal{H}^1$ and its basis states  read $\{ | j \rangle | j \in \tilde N \}$. This space carries the transitive representation $R^{\Sigma_N : \Sigma^{(N-1,1)}}$ of the symmetric group $\Sigma_N$ with the Young subgroup 
\be
\Sigma^{(N-1,1)} = \Sigma_{N-1} \subset \Sigma_N
\ee
as the stabiliser class. This representation can be decomposed
\be\label{rx10}
R^{\Sigma_N : \Sigma^{(N-1,1)}} = \Delta^{( N )} + \Delta^{( N-1,1 )}
\ee
which defines the irreducible basis of the Schur-Weyl duality in the space $\mathcal{H}^1$
\be\label{baza_nieprzyw}
b_{irr} = \{ |\lambda y \rangle \; | \; \lambda \in  \{ (N), (N-1,1) \},\;\; y \in SYT(\lambda)\}.
\ee
Here $SYT(\lambda)$ denotes the set of all standard Young tableaux of the shape $\lambda$, and states $|\lambda y \rangle$
have the form 
\be\label{rx11}
|\lambda y \rangle = \sum_{j \in \tilde N} 
\langle j |\lambda y \rangle 
|j \rangle,
\ee
where probability amplitudes $\langle j | \lambda y \rangle$ can be calculated using formula \cite{magnonic_qudit}
\be\label{coeff}
\langle j |\lambda y \rangle 
=
\left\{
\begin{array}{ccc}
\frac{1}{\sqrt{N}} &\mbox{ for } & \lambda=(N),\;\; y={\scriptsize \Yvcentermath1{\young(12\cdots N)}},\\
-\frac{1}{\sqrt{(j'-1)j'}} & \mbox{ for } & \lambda=(N-1,1), \;\; 1 \leq j <j',\\
\sqrt{\frac{j'-1}{j'}} & \mbox{ for } & j=j',\\
0 & \mbox{ for } & j > j',\\
\end{array}
\right. 
\ee
where $j'$ stands for the Young tableau
\be\label{rx13}
y_{j'} = \Yvcentermath1{\young(1~\ldots~,\j)}, \;\;\; 2 \leq j' \leq N.
\ee
More generally, we can distinguish here the symmetric part of the state (all magnetic configuration with coefficient $-\frac{1}{\sqrt{(j'-1)j'}}$ or $\frac{1}{\sqrt{N}}$) and the special part (magnetic configuration with coefficient $\sqrt{\frac{j'-1}{j'}}$). This structure is strictly connected with RS algorithm (see Sect. \ref{RSalgorytm}).
An example of such states for a Heisenberg magnet with $N=5$ nodes and one ($r=1$) spin deviation is given in Sec. (\ref{przyklad}).

\section{Entangled graphs}\label{entangled_graph}

Following \cite{plesch1} we introduce an object called the entangled graph responsible for the structure of quantum channels. Entangled graph is a quantum structure, in which each qubit is represented as a vertex, and an edge denotes entanglement between connected qubits.

Let us consider a state of the form (\ref{rx11}) with the density matrix 
\be\label{mac_ges}
\rho = |\lambda y\rangle \langle \lambda y |. 
\ee
The reduced density matrix  
\be\label{rdm1}
R_{jk} = \mbox{Trace}_{\mbox{\scriptsize rest of the qubits}} \; (\rho)
\ee
for an arbitrary pair of qubits $j,k$ of Heisenberg magnet can be constructed by performing a partial trace over the rest of the qubits
\be\label{rdm2}
(R_{jk})_{i_j i_k}^{i_j' i_k'} = \sum_
{\mbox{{\tiny 
\begin{tabular}{c}
\mbox{$i_1 \ldots i_N$} \\
\end{tabular}
}}} 
a(i_1 .. i_j .. i_k .. i_N) a^*(i_1 .. i_j' .. i_k' .. i_N),
\ee
where the sum runs over all permutations of nodes $1 \ldots N$ except for nodes $j$ and $k$ and $a(i_1 .. i_j .. i_k .. i_N)$ are coefficients which follow from Eqs. (\ref{rx11},\ref{coeff}), and $i_j \in \tilde 2$ specifies the state of the $j$-th qubit. 
In general, the reduced density matrix for the nodes $(j,k)$ represents a mixed state.
In order to determine how these qubits are entangled amongst themselves we use the concurrence measure
\be\label{conc}
C= max\{ \sqrt{r_1} - \sqrt{r_2} - \sqrt{r_3} - \sqrt{r_4}, 0 \},
\ee
where $\{r_i : i \in \tilde 4\}$ are the eigenvalues in decreasing order of the matrix  
$
R=R_{jk} \tilde R_{jk},
$
where $\tilde R_{jk}$ is the "spin flipped" density matrix
$
\tilde R_{jk} = (\sigma_y \otimes \sigma_y)[R_{jk}]^T (\sigma_y \otimes \sigma_y)
$
and $T$ denotes the transposition of rows and columns.

Here we concentrate only on bipartite aspects of multipartite entanglement by considering the entanglement between a pair $(j,k)$ of qubits in a particular Schur-Weyl state labeled by quantum number $|\lambda y \rangle$ (cf. \ref{rx11}). 
Since the amount of entanglement required to prepare a state monotonically increases with the concurrence, it seems to be a good measure of the strength of the quantum ''bindings'' between qubits. 
Therefore, we choose the concurrence $C$ as a measure of entanglement, and evaluate it on each pair $(j,k)$ of the chain, for all eigenstates of the system. 
Thus, for each state $|\lambda y\rangle$, we calculate $N(N-1)/2$ different bipartite reduced density operators 
$$
\rho_{jk}, \; j < k, \; j,k \in \tilde N,
$$
with matrix elements given by Eq. (\ref{rdm2}). These operators tell us whether nodes $j$ and $k$ are entangled or not, and how strong this quantum correlation is.  This approach allows us to build entangled graph, which characterise quantum structure of the system in a given state $|\lambda y\rangle$. 
If we draw entangled graphs for all one magnon Schur-Weyl states, it turns out that structure of the entangled graph, i.e. distribution of edges between vertices and the strength of these bindings, is entirely coded in the Young tableau $y$.
More precisely, for tableau $y$ of the shape $(N-1,1)$ 
$$
\includegraphics[width=0.4\columnwidth]{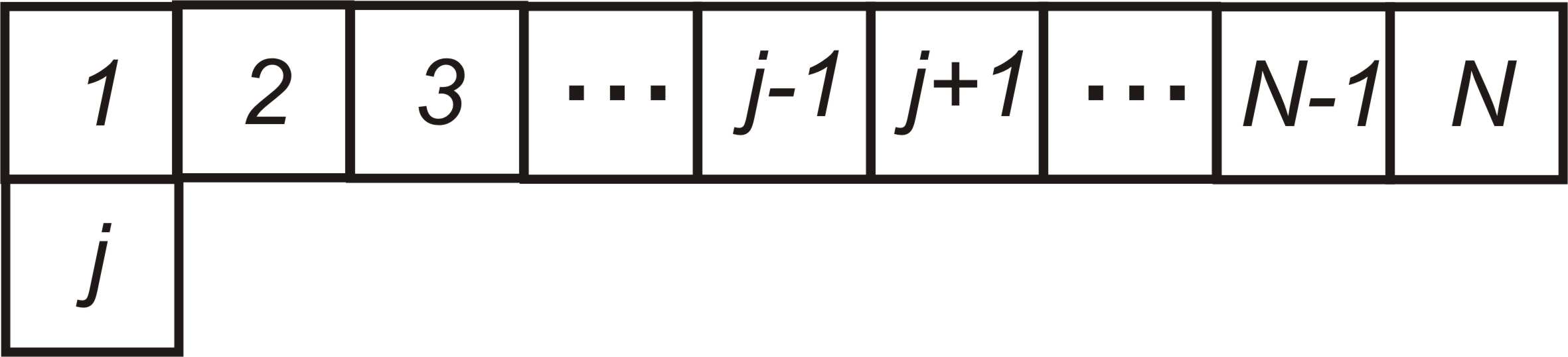}
$$
we have:
\begin{enumerate}
\item The first $j-1$ qubits $1,2, \ldots, j-1$ are entangled peer to peer with concurrence $C_1 = \frac{2}{j(j-1)}$
\item The $j$-th qubit is entangled with all preceding ones $j'=1,2, \ldots, j-1$ with concurrence $C_2 = \frac{2}{j}$
\item All remaining qubits are not entangled with any other.
\end{enumerate}
We observe that the first $j$ qubits of the system form a \emph{complete graph}, in which the last $j$-th qubit is \emph{special}, and all other $N-j$ qubits are absent (cf. \cite{buzek_complete_graf}).
Graphical presentation of this structure is given in Fig. 1.
\begin{figure}[h]
\includegraphics[width=0.5\columnwidth]{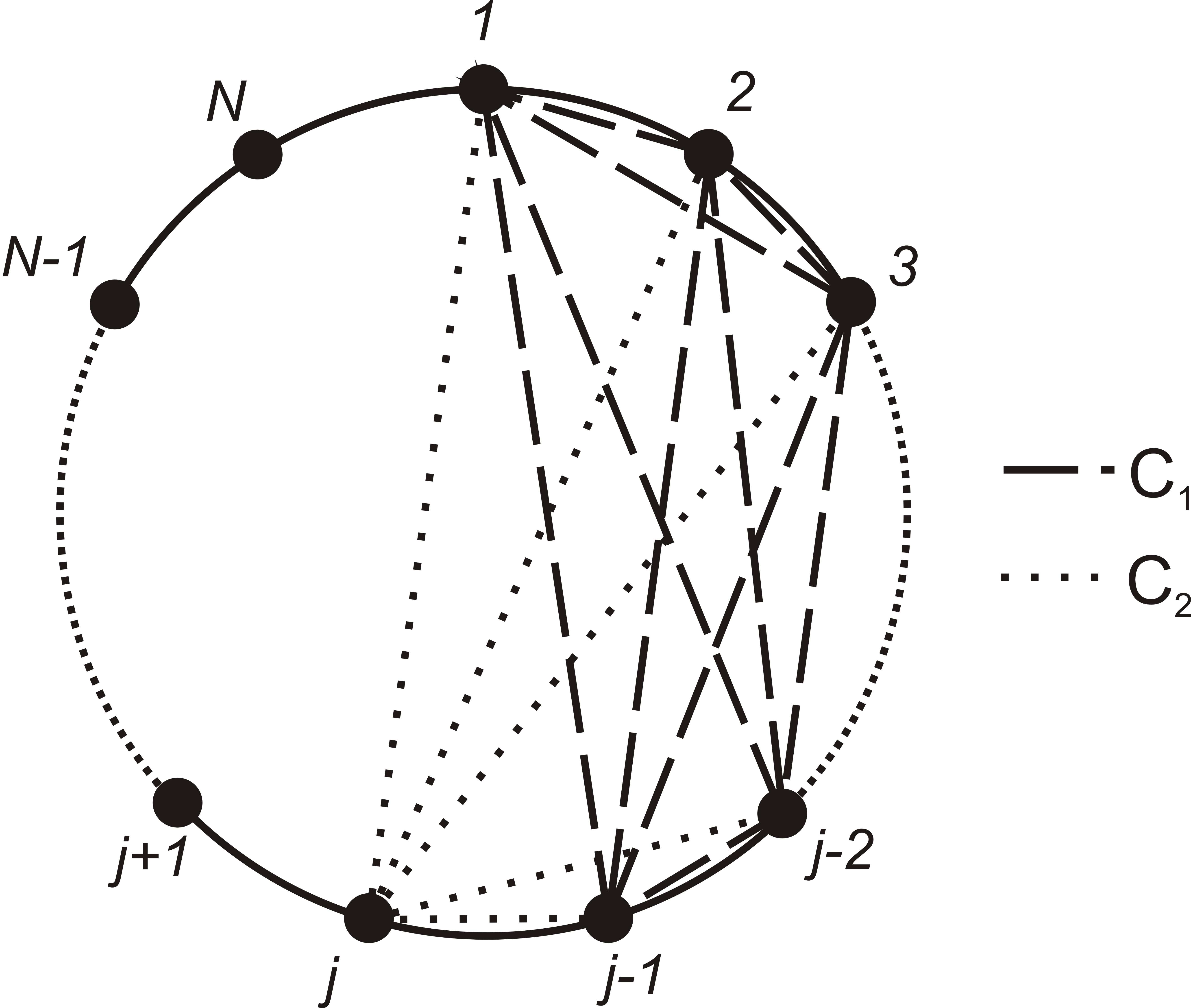}
\caption{Demonstrative picture of the structure of entanglement for the Heisenberg magnet with $N$ nodes and one spin deviation, prepared in the state given by tableau $y$ of the shape $(N-1,1)$. Here we see that all $j$ nodes form a complete graph, out of which the $j$-th node is special. All remaining $N-j$ nodes are not entangled. All $j-1$ nodes are mutually bounded with concurrence $C_1$ while the special node $j$ is bounded with all $j-1$ nodes with concurrence $C_2$.}
\label{fig:1}       
\end{figure}
In addition, for tableau of the shape $(N)$ 
$$
\includegraphics[width=0.4\columnwidth]{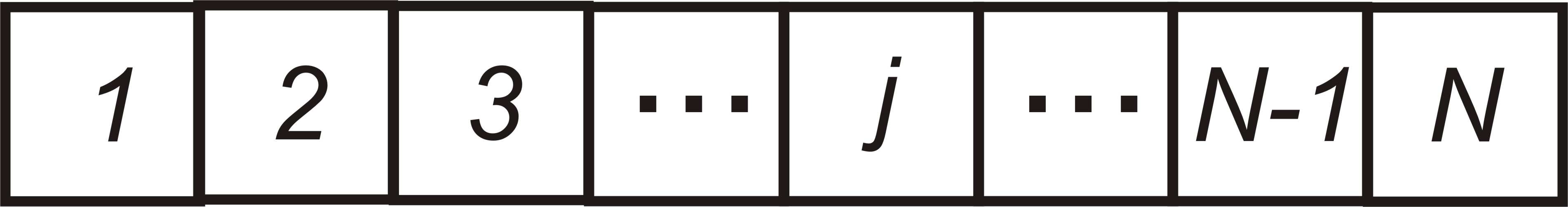}
$$
all $N$ qubits are equally entangled, peer to peer, with concurrence $C = \frac{2}{N}$. See Fig. 2 for the graphical presentation.
\begin{figure}[h]
\includegraphics[width=0.5\columnwidth]{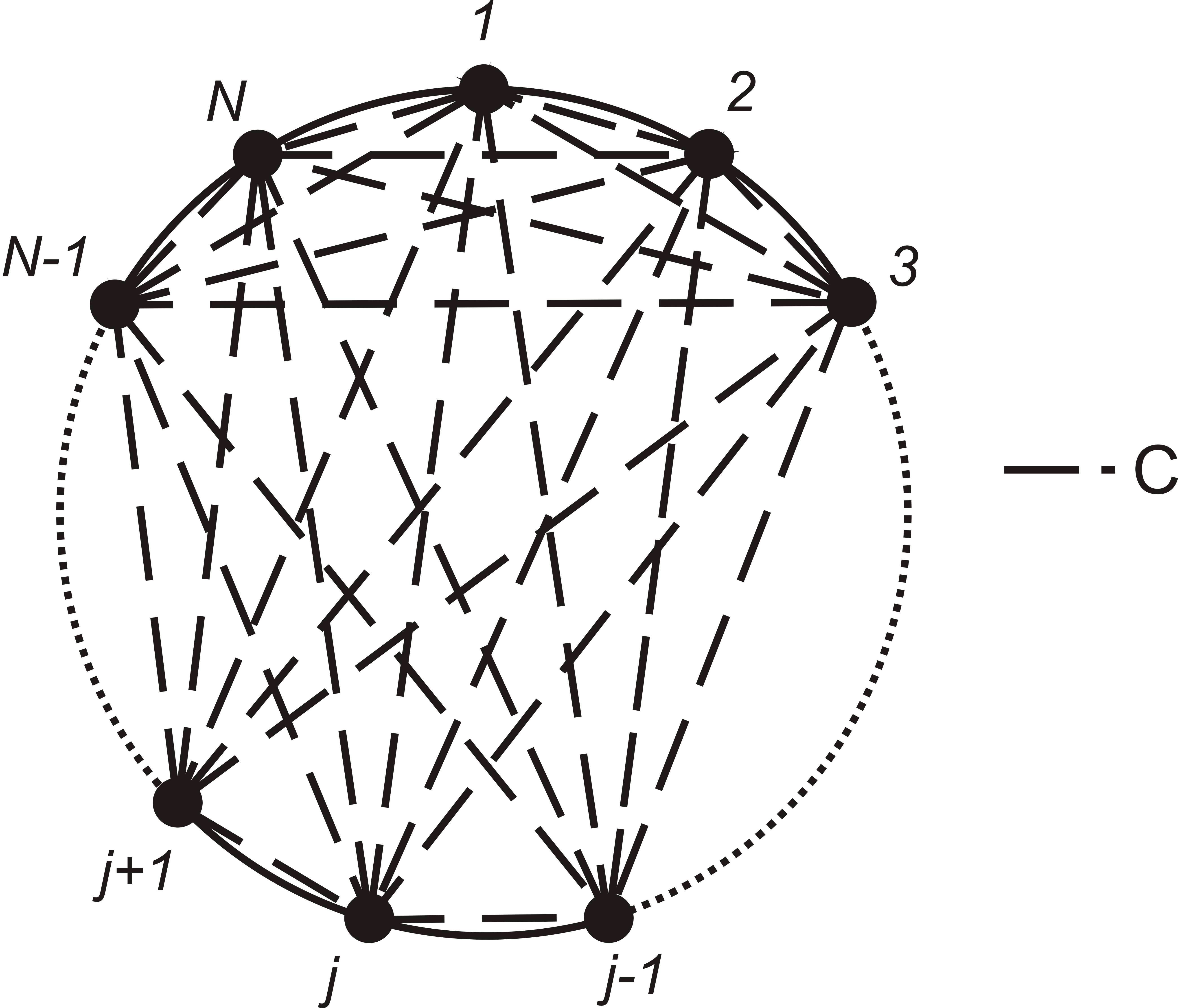}
\caption{Demonstrative picture of the structure of entanglement for the Heisenberg magnet with $N$ nodes and one spin deviation prepared in the state given by tableau $y$ of the shape $(N)$. Here we see that all $N$ nodes form a complete graph i.e. all $N$ nodes are mutually bounded with concurrence $C$.}
\label{fig:2}       
\end{figure}

An example of entangled graphs for the Heisenberg magnet with $N=5$ nodes and one spin deviation is given in Sect. (\ref{przyklad}).

\section{Combinatorial interpretation of entangled graphs} \label{RSalgorytm}

Structure of entangled graphs described in Sec. (\ref{entangled_graph}) can be understood in terms of Young tableau $y$ interpreted in the language of Robinson-Schensted algorithm  \cite{robinson,schensted}. In general settings the RS algorithm is a bijection between magnetic configurations and appropriate irreducible bases of Schur-Weyl duality.
More precisely, let us take the set of words
\be
\tilde{n}^{*}= \bigcup_{N=0}^{\infty} \tilde{n}^{\tilde{N}},
\ee
i.e. the free monoid \cite{monoid_plaktyczny} with juxtaposition of words as monomial multiplication and the empty word $\emptyset$ as the unit. Physically, the free monoid is the disjoint union of sets $\tilde{n}^{\tilde{N}}$ of all magnetic configurations for the rings $\tilde N$, $N=0, 1, \ldots$, with $N=0$ corresponding to the empty set. Let
\be
b^{*}=\bigcup_{N=0}^{\infty} b_{irr}(N)
\ee
be the disjoint union of irreducible bases of the duality of Weyl for the rings $\tilde N$, $N=0, 1, \ldots$, with $b_{irr}(0)=\emptyset$.
The RS algorithm defines a bijection
$
RS: \tilde n^*  \rightarrow b^*
$
between magnetic configurations and appropriate irreducible bases of the Schur-Weyl duality, by putting
\be
RS(f) = (P(f), Q(f)), \,\,\, f \in \tilde n^*,
\ee
where $P(f)$ and $Q(f)$  denotes the Weyl and Young tableau, respectively \cite{sagan}.

In our case this is the bijection between magnetic configurations (\ref{baza}) and Schur-Weyl basis (\ref{baza_nieprzyw}).
The construction of these tableaux for a magnetic configuration
\be\label{konf_mag}
\begin{array}{ccccc}
f=(0 & \ldots & 1 &\ldots & 0)\\
   &        & \uparrow& &\\
   &        & \mbox{$(j-1)$ - th place}& &\\    
\end{array}
\ee
presented in two line notation as
$$ 
\left(
\begin{array}{ccccccc}
1 & \ldots & j-2 & j-1 & j &  \ldots & N\\
0 & \ldots & 0   & 1   & 0 &  \ldots & 0\\
\end{array}
\right )
$$
runs as follows. The first $j-2$ steps yield
\be\label{rs1}
(P_{j'}, Q_{j'}), \; P_{j'}=\Yvcentermath1 \young(0\ldots 0), \; Q_{j'}=\Yvcentermath1 \young(1\ldots \j), \; j'=1, \ldots, j-2,
\ee
single-row tableaux of the shape $(j')$, with $P_{j'}$ having zeros, then at the $(j-1)$-th step
\be\label{rs2}
(P_{j-1}, Q_{j-1})= 
\left (\; 
\begin{array}{|c|c|c|}
\hline
 0 & \ldots & 1 \\
\hline
\end{array}
\; , \;
\begin{array}{|c|c|c|}
\hline
 1 & \ldots & j-1\\
\hline
\end{array}
\; \right )
\ee
the tableau $P_{j-1}$ reaches a single "$1$" at the end, the next, $j$ step produces "a bump", i.e.
\be\label{rs3}
(P_{j}, Q_{j})= \left (\; 
\begin{array}{|c|c|c|}
\hline
 0 & \ldots & 0 \\
\hline
1\\
\cline{1-1}
\end{array}
\; , \;
\begin{array}{|c|c|c|c|}
\hline
 1 & \ldots & j-1 \\
\hline
j\\
\cline{1-1}
\end{array}
\; \right )
\ee
consisting in changing the shape of both tableaux into $(j-1,1)$ interpreted such that the first "0" to the right of "1" in $f$ given by Eq. (\ref{konf_mag}) "bumps" the "1" to the second row of $P_{j}$. The consecutive steps $j'>j$ just enlarge both tableaux by one box in the first row, ending with
$(P_N, Q_N)=(t,y)$.

For example, if we take a magnetic configuration $f=|3\rangle \equiv |00100\rangle$ (c.f. Sect. \ref{przyklad}) we have in two line notation ${12345 \choose 00100}$ and the successive steps of the RS algorithm are the following
$$
{\scriptsize
\begin{array}{ll}
|\emptyset\rangle =  (\emptyset, \emptyset) \stackrel{0}{\longrightarrow}
(\Yvcentermath1 \young(0), \Yvcentermath1 \young(1)) \stackrel{0}{\longrightarrow}
 \left(\Yvcentermath1 \young(00), \Yvcentermath1 \young(12) \right ) \stackrel{1}{\longrightarrow}
 \left(\Yvcentermath1 \young(001), \Yvcentermath1 \young(123) \right ) \\ \\
\stackrel{0}{\longrightarrow}  \left(\Yvcentermath1 \young(000,1), \Yvcentermath1 \young(123,4) \right ) \stackrel{0}{\longrightarrow}  \left(\Yvcentermath1 \young(0000,1), \Yvcentermath1 \young(1235,4) \right ).
\end{array}
}
$$

This RS algorithm gives us a complete insight in the structure of Schur-Weyl states. 
Inserting the first $j-1$ letters (cf. \ref{rs2}) produces the symmetric part of the Schur-Weyl state (what manifest itself in entanglement of $j-1$ qubits of the system with concurrence $C_1$) while inserting the $j$-th letter (cf. \ref{rs3}) gives additional special part of the state (what is responsible for joining the $j$-th qubit to the system, with the concurrence $C_2$).
All remaining $N-j$ steps of RS algorithm have no influence on the structure of entanglement of the system.

Exploiting the RS algorithm we propose a strictly combinatorial algorithm which gives us all entangled graphs for a system, namely:
\begin{enumerate}
\item We take the set of all magnetic configurations\\ $\{|j\rangle~:~j~\in~\tilde N\}$
\item For each magnetic configuration (\ref{baza}) we perform the RS algorithm resulting in irreducible basis (\ref{baza_nieprzyw}).
\item For every elements of the irreducible basis, labelled by the tableau $y$, we can easily draw the entangled graph.
\end{enumerate}
In this way we obtain complete classification of entanglement structure in the given system.

\section{Example}\label{przyklad}

Let us consider an example a Heisenberg magnet with $N=5$ nodes and $r=1$ spin deviation. According to (\ref{rx11}, \ref{coeff}) we have five Schur-Weyl states:
\be\label{przyklad_stanow}
\begin{array}{l}
\left | \; {\scriptsize \Yvcentermath1{\young(1345,2)}}\; \right\rangle= \frac{\sqrt{2}}{2}\left( | 2 \rangle \right)- \frac{\sqrt{2}}{2}\left( | 1 \rangle  \right)\\
\\
\left | \; {\scriptsize \Yvcentermath1{\young(1245,3)}}\; \right\rangle= \frac{\sqrt{6}}{3}\left(| 3 \rangle \right) - \frac{\sqrt{6}}{6}\left(| 1 \rangle + | 2 \rangle \right)\\
\\
\left | \; {\scriptsize \Yvcentermath1{\young(1235,4)}}\; \right\rangle= \frac{\sqrt{3}}{2}\left(| 4 \rangle \right) - \frac{\sqrt{3}}{6}\left(| 1 \rangle + | 2 \rangle + | 3 \rangle \right)\\
\\
\left | \; {\scriptsize \Yvcentermath1{\young(1234,5)}}\; \right\rangle= \frac{2\sqrt{5}}{5}\left(| 5 \rangle \right) - \frac{\sqrt{5}}{10}\left(| 1 \rangle + | 2 \rangle + | 3 \rangle + | 4 \rangle \right)\\
\\
\left | \; {\scriptsize \Yvcentermath1{\young(12345)}}\; \right\rangle = \frac{\sqrt{5}}{5}\left(| 1 \rangle + | 2 \rangle + | 3 \rangle + | 4 \rangle + | 5 \rangle  \right)\\
\end{array}
\ee

These states impose on the system a specific structure of entanglement entirely coded in Young tableaux. Such structure can be presented graphically by use of entangled graphs. The example of entangled graphs for states given in (\ref{przyklad_stanow}) is presented in Table \ref{przyklad_grafow}.

\begin{table}
\caption{The example of entangled graphs for Heisenberg magnet with $N=5$ nodes and one spin deviation. 
The first column presents Schur-Weyl states, the second appropriate entangled graphs while third the values of concurrence.
In the picture of entangled graph the solid line represent a bond with the concurrence $C_1$ while dotted line with $C_2$.}
\label{przyklad_grafow}       
$
\begin{array}{c|c|c}
\hline
\mbox{The Young} & \mbox{The entangled}& \mbox{Concurrence}\\
\mbox{tableau} & \mbox{graph} & {\scriptsize C_1 - \mbox{solid line}, C_2 - \mbox{dotted line}}\\
\hline
&&\\
\left | \; {\scriptsize \Yvcentermath1{\young(1345,2)}}\; \right\rangle & \raisebox{-15pt}{\includegraphics[width=0.2\columnwidth]{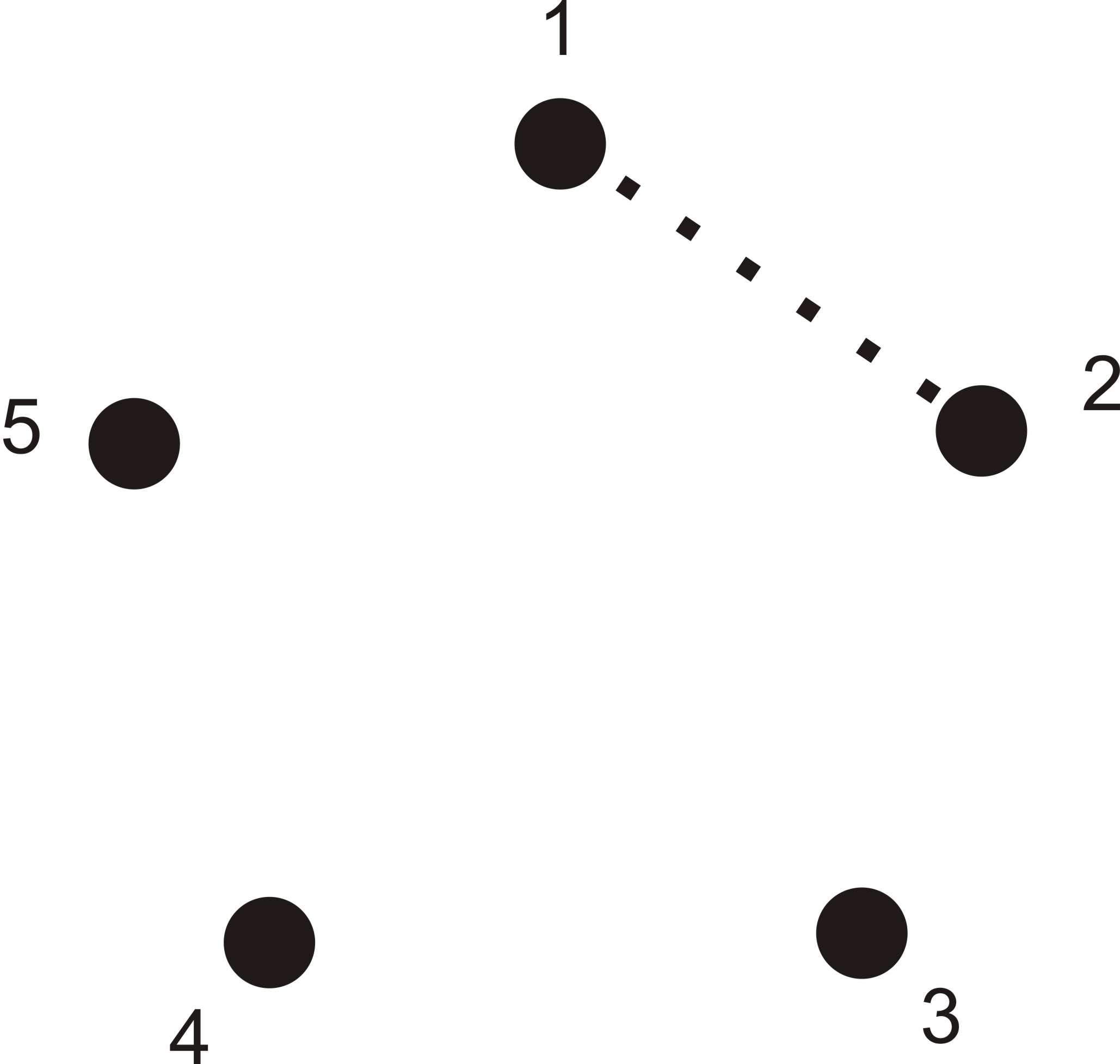}} & C_2=1\\
&&\\
\hline
&&\\
\left | \; {\scriptsize \Yvcentermath1{\young(1245,3)}}\; \right\rangle  & \raisebox{-20pt}{\includegraphics[width=0.2\columnwidth]{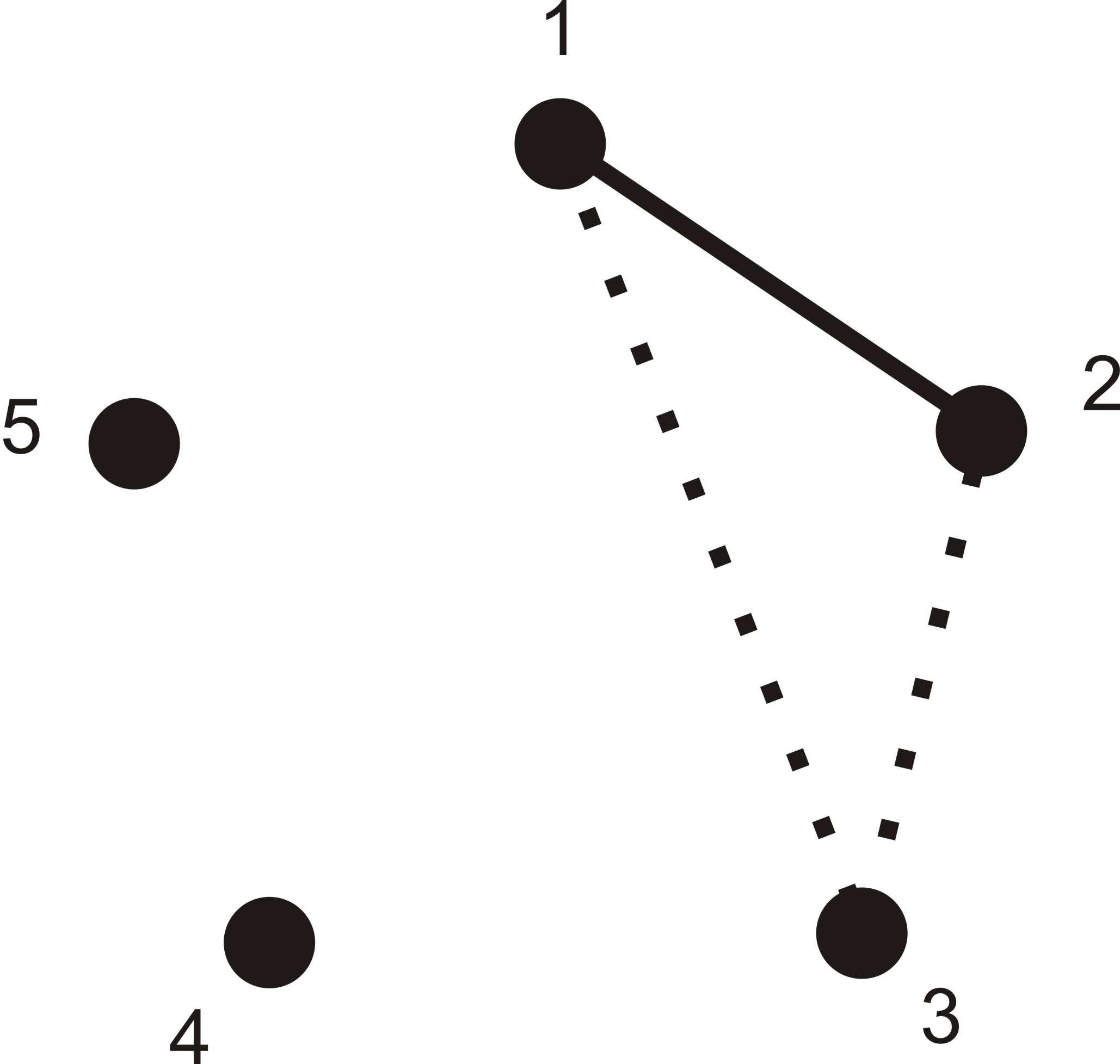}}& C_1=\frac{1}{3},\; C_2=\frac{2}{3}\\
&&\\
\hline
&&\\
\left | \; {\scriptsize \Yvcentermath1{\young(1235,4)}}\; \right\rangle & \raisebox{-20pt}{\includegraphics[width=0.2\columnwidth]{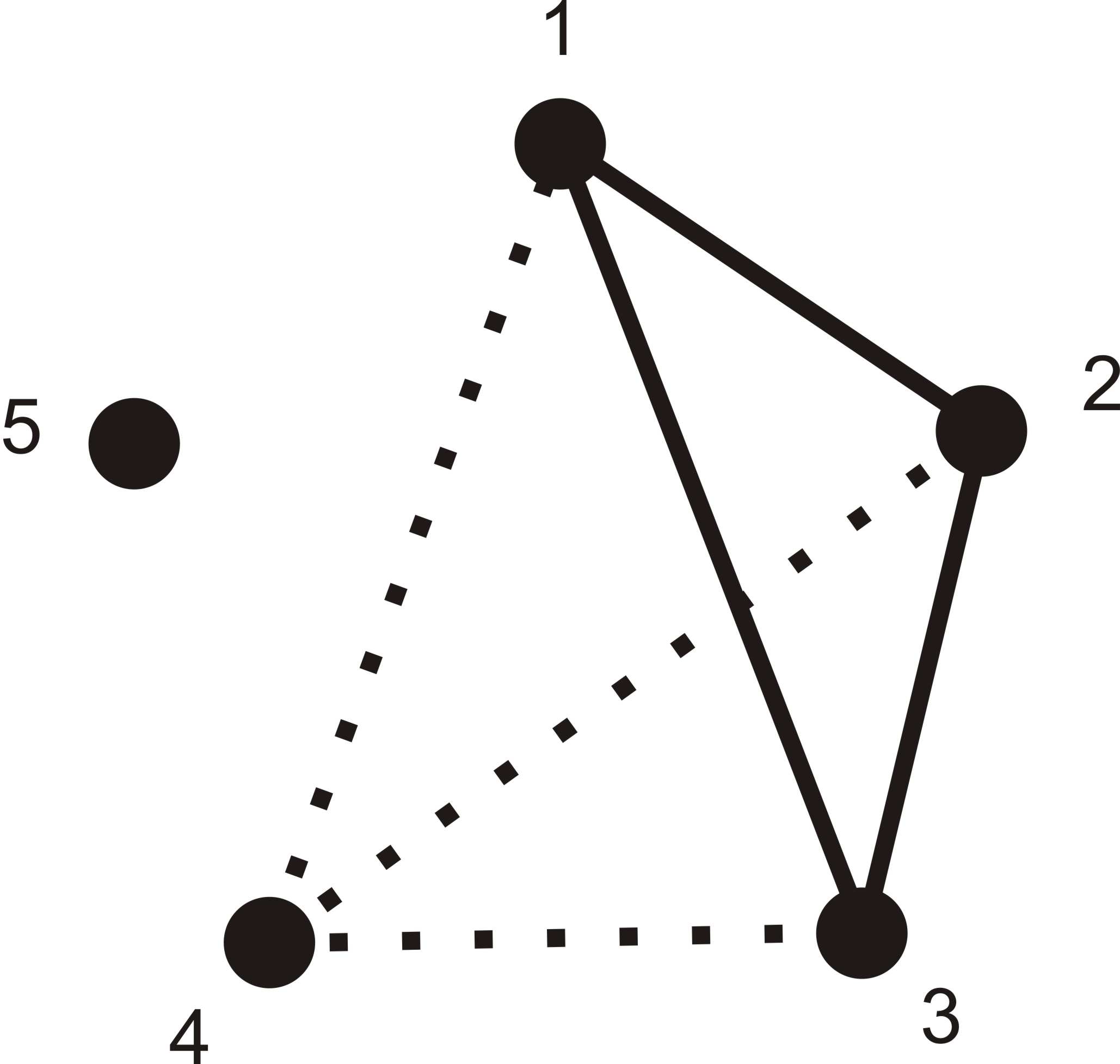}} & C_1=\frac{1}{6},\; C_2=\frac{1}{2}\\
&&\\
\hline
&&\\
\left | \; {\scriptsize \Yvcentermath1{\young(1234,5)}}\; \right\rangle & \raisebox{-20pt}{\includegraphics[width=0.2\columnwidth]{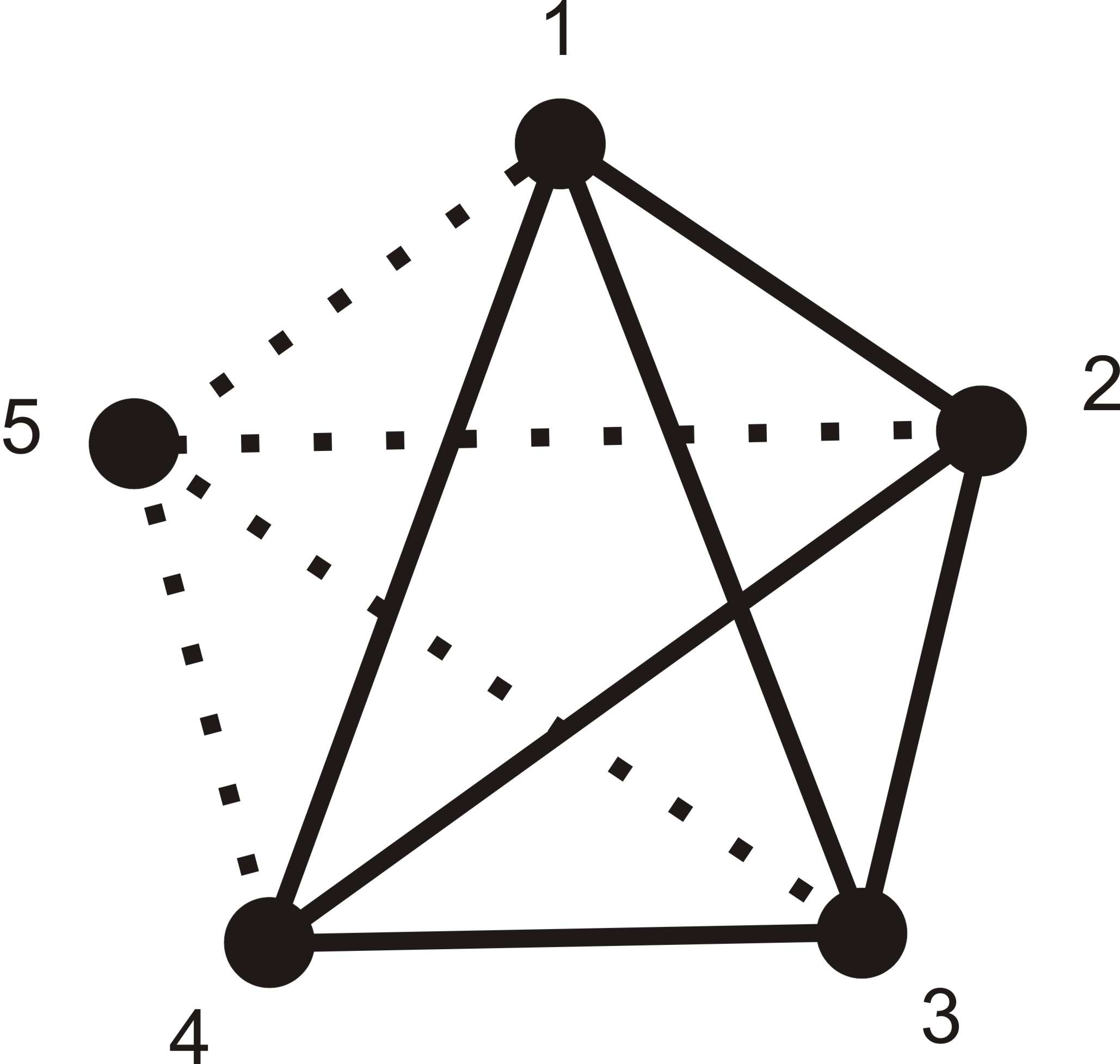}} & C_1=\frac{1}{10},\; C_2=\frac{4}{10}\\
&&\\
\hline
&&\\
\left | \; {\scriptsize \Yvcentermath1{\young(12345)}}\; \right\rangle & \raisebox{-20pt}{\includegraphics[width=0.2\columnwidth]{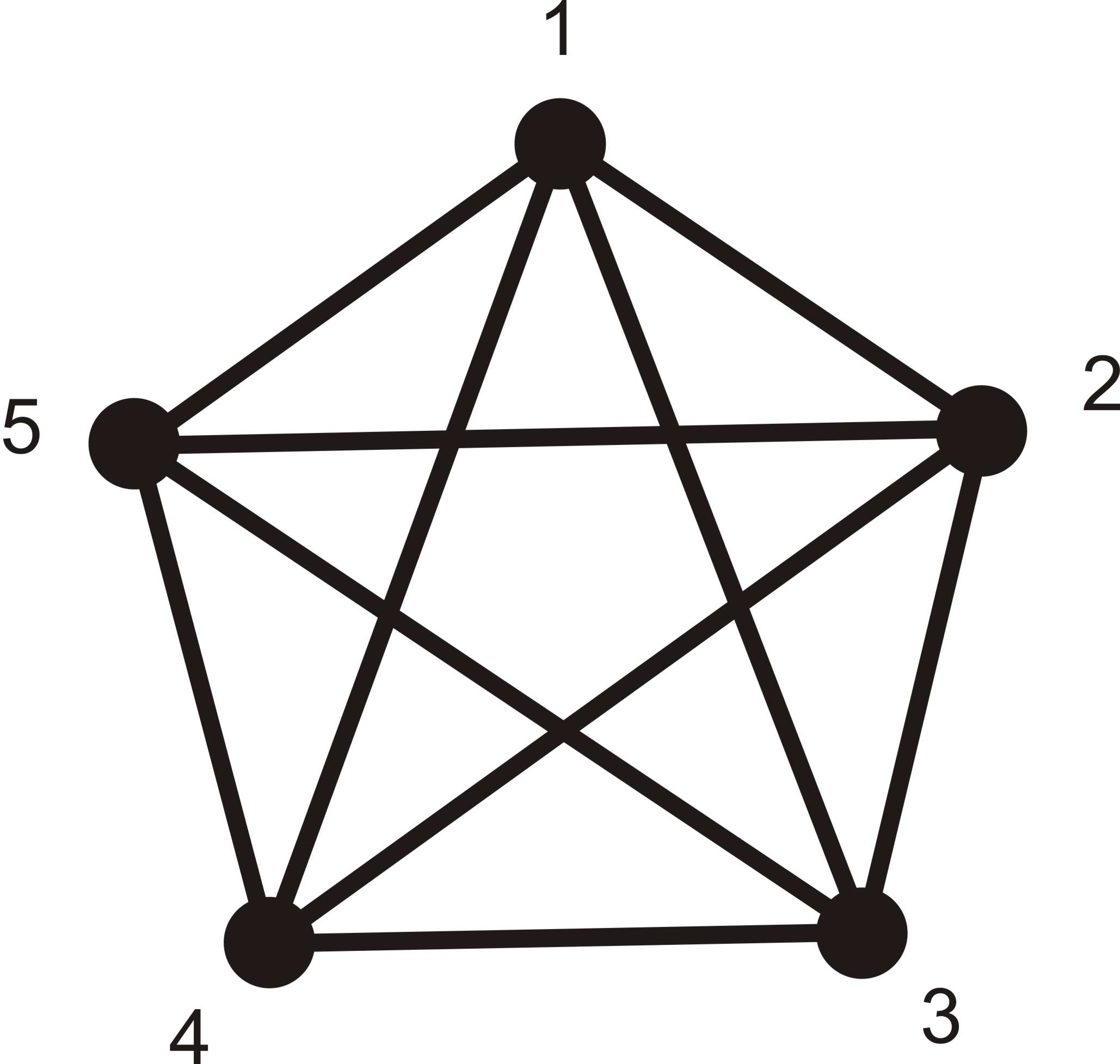}}& C=\frac{4}{10}\\
&&\\
\hline
\end{array}
$
\end{table}

\section{Conclusions}

We have considered a special class of states so called one magnon Schur-Weyl states within the Heisenberg model of $N$ nodes and one spin deviation. We have shown that such states are labelled by Young tableaux and impose on the system the specific structure of entanglement, called entangled graph. 
We also have shown that the structure of entangled graph is entirely coded in Young tableaux what can be explained by use of RS algorithm.
Our results can be used in physical methods for extracting information or resources from a quantum systems, for example spectrum estimation of a density operator \cite{applic_1} or encode quantum information into noiseless subsystems which arise due to collective decoherence \cite{applic_2}.

%

\end{document}